\documentclass[journal]{IEEEtran}

\usepackage{amsmath,amsfonts}
\usepackage{algorithmic}
\usepackage{algorithm}
\usepackage{array}
\usepackage[caption=false,font=normalsize,labelfont=sf,textfont=sf]{subfig}
\usepackage{textcomp}
\usepackage{stfloats}
\usepackage{url}
\usepackage{verbatim}
\usepackage[table]{xcolor}
\usepackage{graphicx}
\usepackage{cite}
\usepackage{xcolor}
\usepackage{multirow}
\usepackage{adjustbox}
\usepackage{rotating}
\usepackage{pdflscape}
\usepackage{booktabs}
\usepackage{hyperref} 
\usepackage{listings}
\usepackage{xcolor}
\definecolor{codegreen}{rgb}{0,0.6,0}
\definecolor{codegray}{rgb}{0.5,0.5,0.5}
\definecolor{codepurple}{rgb}{0.58,0,0.82}
\definecolor{backcolour}{rgb}{0.95,0.95,0.92}

\lstdefinestyle{mystyle}{
    backgroundcolor=\color{backcolour},   
    commentstyle=\color{codegreen},
    keywordstyle=\color{magenta},
    numberstyle=\tiny\color{codegray},
    stringstyle=\color{codepurple},
    basicstyle=\ttfamily\footnotesize,
    breakatwhitespace=false,         
    breaklines=true,                 
    captionpos=b,                    
    keepspaces=true,                 
    numbers=left,                    
    numbersep=5pt,                  
    showspaces=false,                
    showstringspaces=false,
    showtabs=false,                  
    tabsize=2
}
\begin{document}
\bstctlcite{IEEEexample:BSTcontrol}
    \title{Converter: A CEAML Reasoner Python package to Streamline Orchestration Across Cloud and Edge Continuum}
 \author{Ioannis~Korontanis
         Antonios~Makris,
        and~Konstantinos~Tserpes

\IEEEcompsocitemizethanks{\IEEEcompsocthanksitem I. Korontanis, A. Makris and K.Tserpes are with the Department of Informatics and Telematics, Harokopio University of Athens, Greece and also with the School of Electrical and Computer Engineering, National Technical University of Athens. E-mail: \{gkorod,amakris,tserpes\}@hua.gr. 
}
\thanks{}
}

\maketitle

\begin{abstract}

In recent years, there has been a concerted effort in both industry and research sectors to innovate new approaches to DevOps. The primary aim is to facilitate developers in transitioning their applications to Cloud or Edge platforms utilizing Docker or Kubernetes. This paper presents a tool called Converter, designed to interpret a TOSCA extension called CEAML and convert the descriptions into Kubernetes or Kubevirt definition files. Converter is available as a Python package and is recommended for use by orchestrators as an auxiliary tool for implementing CEAML.
\end{abstract}


\IEEEpeerreviewmaketitle

\section{Introduction}
\label{introduction}

Numerous efforts have been documented in the literature aiming to create or extend cloud modeling languages capable of describing the behavior of deployed microservices within a platform.
Supporting tools are often provided alongside the languages to facilitate their implementation.
Achilleos et al. as mentioned in \cite{achilleos2019cloud}, refer to tools that convert high-level cloud modeling language descriptions into technical ones as reasoners.

A cloud modeling language that is quite popular in the research field is TOSCA\footnote{\url{https://docs.oasis-open.org/tosca/TOSCA-Simple-Profile-YAML/v1.3/os/TOSCA-Simple-Profile-YAML-v1.3-os.html}}, developed by OASIS. 
TOSCA aims to depict how multi-component applications are deployed across Cloud or Edge resources in a way that's not tied to any specific technology. Many papers discuss methods for utilizing TOSCA or its extensions, in conjunction with a reasoning tool, to facilitate application migration. For example, Tosker\cite{Toskerarticle} is a tool capable to translate models written with a TOSCA extension, delineating Docker containers within Docker deployment plans.
Another case is 
CloudCAMP\cite{CloudCampinproceedings}, which translates TOSCA into infrastructure as code, which serves as deployment plans for the described applications in the presented platform. A more sophisticated tool named MICADO Submitter is presented by DesLaurier et al. \cite{deslauriers2021cloud}. MICADO Submitter is capable to translate descriptions written in a TOSCA extended language into deployment plans to Docker Swarm and Kubernetes. This tool was utilized by the orchestrator of the system which applied the deployment plans. TOSCA can also facilitate deployments that aren't tied to containerization. A notable example is the ARIA project, detailed at \footnote{\url{https://incubator.apache.org/projects/ariatosca.html}}, which is able to convert TOSCA models into deployments for applications across various cloud providers such as Microsoft Azure Cloud, Amazon Web Services, VMware, and OpenStack.

Similarly, to streamline the implementation of our cloud modeling language, CEAML, we developed a reasoner called Converter.
CEAML is designed based on the research conducted on \cite{app14062311}. It's main purpose is to describe the deployment and runtime adaptation of applications in Cloud and Edge environments. CEAML is an extension of TOSCA designed to employ a more user-friendy syntax for describing application components, resources and orchestrator behaviour based on event-driven logic.
This paper does not delve into the semantics of the language; rather, it offers guidance on utilizing Converter to implement it within a system.
It's important to clarify that the Converter alone does not function as an orchestrator. However, it can be employed by an orchestrator to execute deployment and runtime adjustments according to the descriptions outlined in a CEAML model.
The outputs of the Converter are configuration files for Kubernetes or Kubevirt. Kubevirt\footnote{\url{https://kubevirt.io/}} is a plugin that could be installed on Kubernetes to enable the deployment of virtual machines. 

The rest of the paper is organized as follows:
\begin{itemize}
\item Section \ref{pack_function} delves into the functionalities Converter provides to an orchestrator for executing specified tasks.
\item Section \ref{impact} investigates the importance of integrating Converter as a reasoner and introduces a proposed pipeline
\item Section \ref{conclusion} summarizes the conclusions drawn from the utilization of the Converter reasoner.
\end{itemize}

\section{Package Functionalities}
\label{pack_function}
\begin{figure}[htbp]
    \centering
    \includegraphics[width=0.45\textwidth]{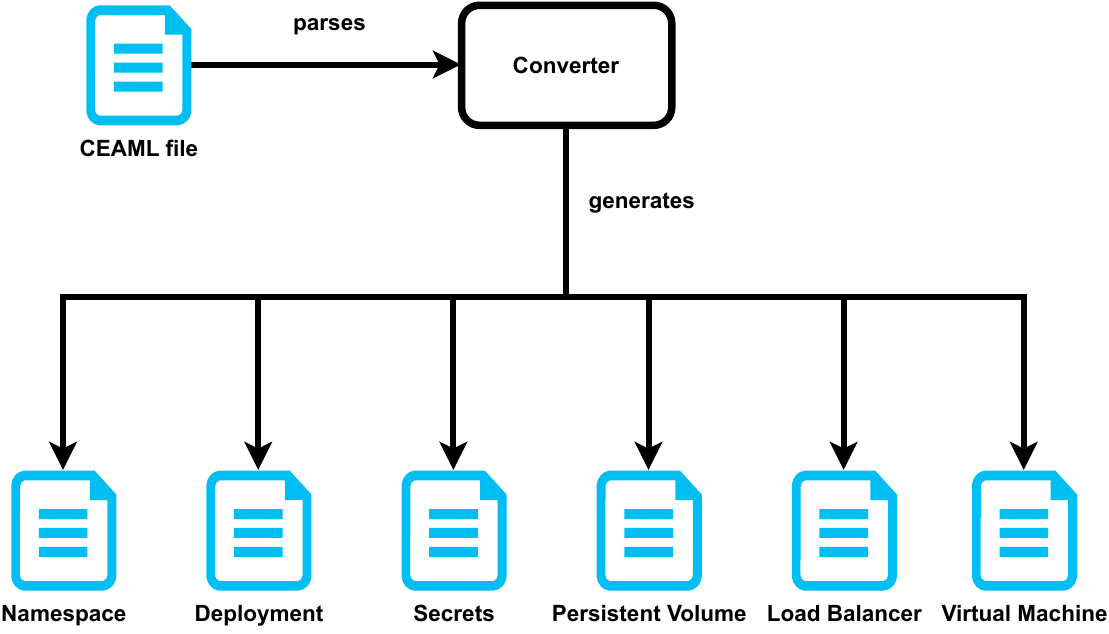} 
    \caption{Converter input and outputs}
    \label{fig:converter}
\end{figure}

Converter is provided as a Python package\footnote{\url{https://pypi.org/project/converter-package/}} and its source code is available on Github\footnote{\url{https://github.com/Efficient-Computing-Lab/Converter}}.  
To generate the necessary outputs, Converter requires the appropriate CEAML model (Figure \ref{fig:converter}).
For an application described in CEAML, Converter is able to generate multiple Kubernetes and Kubevirt configurations:
\begin{itemize}
    \item A namespace with a unique ID that would contain application components. The unique ID is required to identify components that belong to the same application and being deployed into different clusters.
    \item Multiple deployment plans for Kubernetes, one for each application component. These deployment plans are required for components that should be deployed as pods.
    \item Multiple secrets, one for each application component. Secrets are needed for Kubernetes and Kubevirt in order to pull Docker or virtual machine images from private registries.
    \item Multiple persistent volumes for application components that need to have persistent storage
    \item Multiple load balancers for application components that need to have an external IP
    \item Multiple virtual machine deployments for each application component that need to be deployed as a virtual machine on Kubevirt.
\end{itemize} 

From the aforementioned outputs and their variations, the Converter facilitates three primary orchestration operations. These operations are applicable to either a single Kubernetes or Kubevirt cluster, or across multiple clusters:

\begin{itemize}
\item Generate deployment plans for application components on Kubernetes or Kubevirt (Section \ref{deployment}).
\item Generate termination plans for application components that run on Kubernetes or Kubevirt clusters (Section \ref{termination}).
\item Generate scale out plans for application components that run on Kubernetes or Kubevirt clusters (Section \ref{scale out}). 
\end{itemize}

\subsection{Deployment Plans}
\label{deployment}
To generate a deployment plan for a CEAML-described application on Kubernetes or Kubevirt, developers must provide the following necessary inputs to the Converter:

\begin{enumerate}
\item Provide access tokens for the private registry housing their images and encode them in base64. 
\item Specify the path of the model written in CEAML
\item Indicate the version of their application 
\item Provide external IPs that could potentially be used by a component
\item Specify the cluster ID to which the application should be deployed
\item Provide a list of available GPUs on the cluster. If there are no GPUs available, provide an empty list.
\end{enumerate}

\begin{figure}[htbp]
    \centering
    \includegraphics[width=0.45\textwidth]{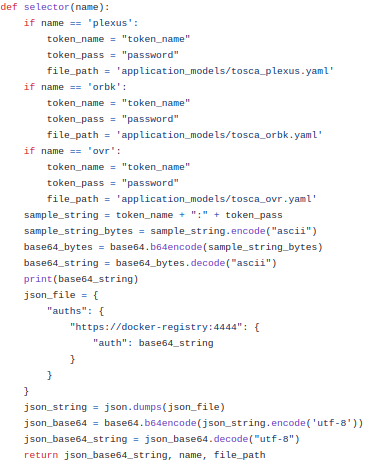} 
    \caption{Code snippet to provide access tokens and CEAML models}
    \label{fig:selector_fig}
\end{figure}

An example demonstrating how to provide inputs 1. and 2. is depicted in Figure \ref{fig:selector_fig}. In this scenario, there's a function called ``selector'' that requires the application's name as input. If the name matches one of the predefined ones, it retrieves the access tokens and creates the base64-encoded version of the access token. Additionally, the function defines the path of the model written in CEAML. Finally, the function outputs the base64 access token, the application's name, and the model's path.

\begin{figure}[htbp]
    \centering
    \includegraphics[width=0.45\textwidth]{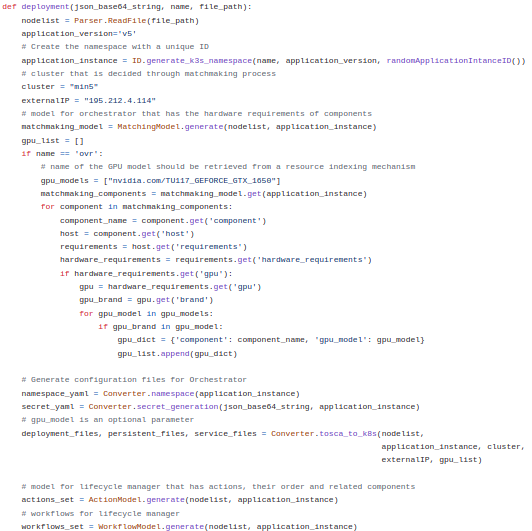} 
    \caption{Code snippet for deployment plans}
    \label{fig:deployment_fig}
\end{figure}

Figure \ref{fig:deployment_fig} presents a way to provide inputs 3., 4., 5. and 6. The code snippet presents a function named deployment that retrieves the outputs of the previously mentioned selector function. The application version, cluster ID, external IP and GPU list are hardcoded in order to simplify the example.
In any situation, you can also input these values through a mechanism instead of embedding them directly into your code.

Combining the aforementioned figures provides a comprehensive example of generating deployment files with Converter.
Figure \ref{fig:deployment_fig} demonstrates that Converter requires several functions to be called for generating deployment plans.
The explanation of these functions are presented in Table \ref{tab:deployment_functions}.

\begin{table}[htbp]
\centering
\small
\caption{Description of deployment related functions}
\label{tab:deployment_functions}
\begin{tabular}{p{0.40\linewidth}|p{0.5\linewidth}}
\hline
\textbf{Function} & \textbf{Description} \\
\hline
\textit{ID.generate\_k3s\_namespace, Converter.namespace} & Generate a namespace with a unique ID. The ID.generate\_k3s\_namespace function requires the application name, version, and a random alphanumeric. After generating the unique ID, call Converter.namespace and pass the unique ID as a parameter to create the K3s namespace. (Required) \\
\hline
\textit{MatchingModel.generate} & Generate a model for the orchestrator with hardware requirements of components. This function requires a list containing all instances of CEAML entities (nodeList) and the unique ID of the instance to be deployed (application\_instance). (Optional) \\
\hline
\textit{Converter.secret\_generation} & Generate required secrets. This function requires the base64 version of access tokens and the unique ID of the instance to be deployed. (Required) \\
\hline
\textit{Converter.tosca\_to\_k8s} & Generate deployment files, persistent volumes, and services. This function requires a list containing all instances of CEAML entities (nodeList), the unique ID of the instance to be deployed (application\_instance), cluster ID, potential external IP, and potential GPU list. (Required) \\
\hline
\textit{ActionModel.generate} & Generate a model that includes components and related actions. This function requires a list containing all instances of CEAML entities (nodeList) and the unique ID of the instance to be deployed (application\_instance). (Optional) \\
\hline
\textit{WorkflowModel.generate} & Generate a model that includes workflows with runtime adaptation actions to be performed under certain conditions. This function requires a list containing all instances of CEAML entities (nodeList) and the unique ID of the instance to be deployed (application\_instance). (Optional) \\
\hline
\end{tabular}
\end{table}

In case developers want to use a conventional deployment on Kubernetes or Kubevirt, they need to execute the required steps.
However, in case they have created a distributed orchestrator that operates across clusters, they may also consider utilizing the optional steps (as indicated in Table \ref{tab:deployment_functions}) to assist in the orchestration process.
Especially beneficial for distributed orchestrators are the matchmaking model and the workflow model.
The matchmaking model will provide a submodel of CEAML that includes the described application components and their hardware requirements, in order to assist orchestrators to find the desirable hosts. On the other hand, the workflow model contains instructions on how the orchestrator should be perform runtime adaptation under certain condition. These models offer developers the option to deploy and scale applications differently from the methods provided by Kubernetes
The outputs of the mentioned functions are stored in memory, giving developers the flexibility to either save them as files or utilize them directly.


\subsection{Termination Plans}
\label{termination}
To enable Converter to generate termination plans for a CEAML-described application on Kubernetes or Kubevirt, developers must provide the following required inputs:

\begin{itemize}
\item The name of the running instance, for example \textit{acc-uc2orbk-0-0-4-00036-gameserver-7reio-min1}
\item The path of the model written in CEAML
\end{itemize}

\begin{figure}[htbp]
    \centering
    \includegraphics[width=0.45\textwidth]{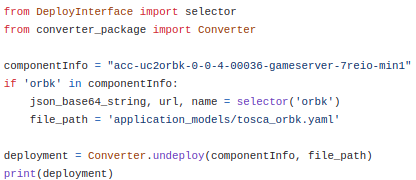} 
    \caption{Code snippet for termination plans}
    \label{fig:termination}
\end{figure}

Similarly to the deployment functions, all required inputs for the termination function can be provided either manually or through a mechanism. A usage example demonstrating how to generate the termination plans with Converter is presented in Figure \ref{fig:termination}.

As illustrated in the above code snippet, only the \lstinline{Converter.undeploy(componentInfo, file_path)} should be invoked. This method internally calls the Parser, which parses the CEAML model and generates the termination plan for the running instance.

\subsection{Scale Out Plans}
\label{scale out}

To enable Converter to generate scale-out plans for a CEAML-described application on Kubernetes or Kubevirt, an orchestrator or another mechanism must retrieve the plans and deploy them with the same running instance ID to a separate cluster. This scenario is exclusive to distributed orchestrators. Developers should provide the following required inputs:

\begin{itemize}
\item The name of the running instance, for example \textit{acc-uc2orbk-0-0-4-00036-gameserver-7reio-min1}
\item The path of the model written in CEAML
\end{itemize}

\begin{figure}[htbp]
    \centering
    \includegraphics[width=0.45\textwidth]{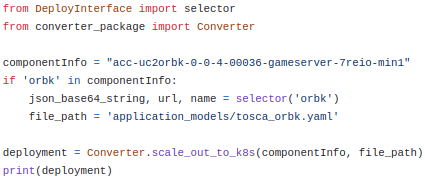} 
    \caption{Code snippet for scale out plans}
    \label{fig:scale out}
\end{figure}

Similarly to the previously mentioned functions, all required inputs for the scale-out function can be provided either manually or through a mechanism. A usage example on how to generate the scale out plans to perform scaling with Converter is presented in Figure \ref{fig:scale out}.

The scale-out function, similar to the termination function, will trigger the Parser in the background. The function will parse the CEAML model and generate the scale-out plan for the running instance.

\section{Impact}
\label{impact}
Utilizing Converter enables platforms to leverage the descriptive capabilities of CEAML language, facilitating deployment and runtime adaptation beyond the functionalities offered by Kubernetes and Kubevirt alone. A distributed orchestrator could employ the Converter in order to perform matchmaking process and deployment of the described application components across different Kubernetes or Kubevirt clusters. 

CEAML and the Converter have already been used by a European project named ACCORDION \cite{10.1145/3452369.3463816}. The distributed orchestrator \cite{carlini2023smartorc} of the ACCORDION platform, managed to carry out all orchestration processes supported by the Converter across Kubernetes and Kubevirt clusters situated on both Edge and Cloud environments. 
The orchestrator achieved a significant milestone by leveraging CEAML and the Converter to autonomously execute deployment and runtime adaptation, without depending on Kubernetes orchestrator and scaling capabilities.

Following that, the Converter can assist in a range operations, spanning from deploying an application to a single Kubernetes cluster to assisting distributed orchestrators in deploying, terminating or scaling across multiple Kubernetes and Kubevirt clusters.
\section{Conclusion}
\label{conclusion}

Converter reasoner seamlessly integrates CEAML orchestration semantics into both simple and complex orchestration mechanisms. Its primary advantage lies in its  ability to generate Kubernetes or Kubevirt configuration plans, streamlining the orchestration process outlined in CEAML. Developers can utilize the Converter's outputs by applying them to a single Kubernetes cluster or even to an orchestration mechanism overseeing multiple clusters.

The Converter can be easily installed as it is distributed as a Python package. Its source code is openly available on a public GitHub repository, which allows the cloud modeling languages community to delve into the code, propose new features, or gain insight into the expansion of TOSCA to develop CEAML.


\ifCLASSOPTIONcaptionsoff
  \newpage
\fi


\vfill


\end{document}